\documentclass{PoS}
\usepackage{ amssymb, fontenc, times, mathptmx, graphicx}
\usepackage{pgfplots,placeins,bm,xcolor,color}
\usepackage{pgfplotstable,rotating,datatool,filecontents,
	tikz,ifthen}
\usepackage[utf8]{inputenc}

\title{Effects of jet-medium interactions on angular correlations of jet-particle pairs at different energy scales}

\ShortTitle{Angular correlations at different energy scales}

\author{\speaker{Martin Rohrmoser}$^a$, Pol Gossiaux$^b$, Thierry Gousset$^b$, J\"org Aichelin$^b$, Iurii Karpenko$^b$\\
        \llap{$^a$}Institute of Physics, Jan Kochanowski University, 25-406 Kielce, Poland
\\
        \llap{$^b$}SUBATECH, IN2P3/CNRS, Universit\'e de Nantes, IMT Atlantique, 4 rue Alfred Kastler, 44307
Nantes Cedex 3, France
\\
 E-mail: \email{rohrmoser.martin1987@gmail.com}, \email{Pol-Bernard.Gossiaux@subatech.in2p3.fr}, \email{gousset@subatech.in2p3.fr}, \email{Aichelin@subatech.in2p3.fr}, \email{yu.karpenko@gmail.com
}        
        }

\abstract{The energy-loss of hard probes within the hot and dense medium of a quark gluon plasma (QGP) can be described by theoretical models based on radiative energy loss as well as combinations of collisional and radiative energy loss. In a search for observables that allow to disentangle these energy-loss mechanisms, we introduced a set of effective models that allows to investigate the consequences on jets of both types of jet-medium interactions within a consistent framework.
We particularly studied angular jet-broadening via angular two-particle correlations.
Distinguishing contributions from pairs at different scales of particle momenta, we found qualitative differences between radiative and collisional approaches.}

\FullConference{International Conference on Hard and Electromagnetic Probes of High-Energy Nuclear Collisions\\
		30 September - 5 October 2018\\
		Aix-Les-Bains, Savoie, France}

\begin{document}

\section{Introduction}
Highly energetic and/or heavy particles (hard probes) traversing the QGP lose part of their energies due to in-medium interactions.
Even though the combinations of the observables of the nuclear modification factor $R_{\rm AA}$ and the elliptic flow $v_2$ put constraints on models for in-medium energy loss, numerous theoretical models based on either radiative energy-loss processes or a combination of radiative and collisional processes are in good agreement with experimental data (cf.~\cite{Andronic2016} for a review). 
Correlations between particle pairs from within jets of hard particles directly access the additionally emitted jet-particles due to processes of radiative energy loss and are, thus, promising tools to resolve part of the ambiguity. 
To this end, we introduced a framework of effective models that allow to consistently study collisional and radiative processes of energy loss, simulated large sets of jets in the corresponding Monte-Carlo algorithms and extracted and studied qualitatively the resulting angular two-particle correlations.
\section{Approach}
In this study jets in the vacuum and the medium are described by a Monte-Carlo simulation of the collinear splittings due to bremsstrahlung, already present in the vacuum, together with possible medium modifications that happen simultaneously to the jet-evolution. Jets are simulated as timelike parton cascades, originating from an initial parton with energy $E_{\rm ini}$ and virtuality $Q_\uparrow$. Cascade particles with a certain virtuality $Q$ above a threshold $Q_\downarrow$ can undergo multiple collinear splittings. The simulation of a single cascade is finished, when every cascade-particle reaches virtuality $Q_\downarrow$.

In the vacuum, jet evolution is given by multiple collinear splittings that follow the Dokshitzer-Gribov-Lipatov-Altarelli-Parisi (DGLAP) equations. 
Thus, the momentum fractions of the splitting products and the virtuality $Q$ of the splitting parton were selected from the leading order (LO) timelike DGLAP splitting functions and the corresponding Sudakov-factors, respectively. The times $\tau$ between two consecutive splittings can be estimated, using the uncertainty principle, via the particle virtuality and its energy $E$   (cf.:~\cite{Zapp:2008gi,Renk2008} for similar approaches) as
\begin{equation}
\tau=\frac{E}{Q^2}\,,
\label{eq:timeest}
\end{equation}
within the rest-frame of the medium.

Medium modifications of jet-evolution were effectively included by changing the four-momenta $p=(E,\vec{p})$ of partons between consecutive splittings continuously over time $t$. We considered three effective models for the description of jet-energy loss: A model for purely collisional energy loss, a model for purely radiative energy loss, and a hybrid model that combines both types of energy loss.

For the description of collisional energy loss, we considered a continuous drag-force $\vec{A}(t)$, anti-parallel to the incoming parton three-momentum $\vec{p}_\|$ and a stochastic transverse momentum transfer $\hat{q}_C$ acting in a random direction $\vec{p}_\perp$ orthogonal to $\vec{p}_\|$
\begin{eqnarray}
\hat{q}_C(t)=\frac{d \langle\vec{p}_\perp\rangle^2}{dt}\,,
&&
\vec{A}(t)=-\frac{d}{dt}\langle \vec{p}_\| \rangle\,.
\label{eq:modcoll}
\end{eqnarray}

We assumed that jet-particles interact with a medium that is already at thermal equilibrium. Therefore, an Einstein-Smoluchowski relation between $\hat{q}_C$ and $\vec{A}$ holds, for which we followed~\cite{Berrehrah:2014kba} and used 
\begin{equation}
\frac{\hat{q}_C}{A}=0.09+0.715\frac{T}{T_c}\,\,\,\,[{\rm GeV}]\,,
\label{eq:qhatAratio}
\end{equation}
where for the $T_c$, the critical temperature of the medium, a value of $0.16$~GeV was assumed.
For the temperature dependence of the transverse momentum transfer, we used the parametrization 
\begin{equation}
\hat{q}_C=\frac{210}{(1+53T)}T^3\,,
\label{eq:parammed}
\end{equation}
 for quarks propagating in the medium. For gluons, the right hand side of Eq.~(\ref{eq:parammed}) is multiplied by a factor $C_A/C_F$. 
For this study the temperatures were obtained from hydrodynamic simulations of the medium using the EPOS3-HQ approach~\cite{Werner:2013tya} for the $10\%$ most central Pb-Pb collisions at $\sqrt{s_{\rm NN}}=2.76$~TeV. For simplicity, we used then a time-dependent temperature profile $T(t)$ in the Monte-Carlo simulations of jets, where we assumed a straight path starting from the center of a medium.

In the model of purely collisional energy loss, the energy $E$ of a jet-particle changes over a time-interval $\Delta t$ as
\begin{eqnarray}
E(t+\Delta t)=\sqrt{E(t)^2+\Delta t((\hat{q}_C(t)-2\|\vec{p}(t)\|A(t)))+\mathcal{O}\left(\Delta t^2\right)}\,.
\label{eq:yajemDEincr}
\end{eqnarray}
Thus, the drag force $\vec{A}(t)$ leads to an energy transfer from jet-particles with $\|\vec{p}\|\gg T$ to the medium, while energy can also be transfered from the medium to the jet-particle if $\|\vec{p}\|<\frac{\hat{q}_C}{2A}$. As the current study focuses on highly energetic jets that do not thermalize, the drag force yields a jet-energy loss.

For the effective simulation of radiative energy-loss, we used the basic assumption of the YaJEM approach~\cite{Renk2008} that during the in-medium propagation of a jet-particle its virtuality $Q$ is increased by a rate $\hat{q}_R$, i.e.:
\begin{equation}
\frac{d }{dt}Q^2=\hat{q}_R(t)\,,
\label{eq:yajemcont}
\end{equation}
where for simplicity we assumed $\hat{q}_R=\hat{q}_C$.

The increase in virtuality can be attributed to either a loss in parton momentum during in-medium propagation, or an increase in energy. 
Since we want to study the effects of additional radiation separately from other effects, we chose the second option.

A model where jet-parton energy is transfered to medium-particles during inelastic scatterings can still be obtained by combining the effective model for radiative energy loss with the model for collisional energy loss in a hybrid approach where the jet-parton four-momenta are changed following both Eq.~(\ref{eq:modcoll}) and Eq.~(\ref{eq:yajemcont}) simultaneously.

The three considered models for jet-medium interactions are summarized in Tab.~\ref{tab1} together with their effects on jet-particle four-momentum components.

	\begin{table}[h!]
		\centering
		\begin{tabular}{l|cccc}
			model &Q&$\vec{p}_\|$&$\vec{p}_\perp$&E\\\hline
{radiative}&$\uparrow$&$=$&$=$&$\uparrow$\\
{collisional}&$=$&$\downarrow$&$\uparrow$&$\pmb{\bm{\downarrow}}\uparrow$\\
{hybrid}&$\uparrow$&$\downarrow$&$\uparrow$&$\pmb{\bm{\downarrow}}\uparrow$	
		\end{tabular}
		\caption{Summary of the effective models of jet-medium interaction: The symbols indicate, whether, between two consecutive splittings, the $4$-momentum components of jet-particles increase ($\uparrow$), decrease ($\downarrow$), remain unchanged ($=$), or decrease for large $\|\vec{p}\|$ and increase for small $\|\vec{p}\|$ ($\pmb{\bm{\downarrow}}\uparrow$).}
		\label{tab1}
	\end{table}
\section{Momentum scale dependence of angular correlations }

In order to better understand the contributions of hard and soft jet-particles to angular jet broadening, we studied the angles $\Delta \theta$ between a leading trigger particle and associated jet-particles as a function of a momentum scale cut $\|\vec{p}\|_{\rm cut}$. For a given jet, jet-particles with three-momenta $\vec{p}$ that do not satisfy $\|\vec{p}\|\geq \|\vec{p}\|_{\rm cut}$ were removed from the jet. For the still possible particle pairs within an individual jet, the angles $\Delta \theta$ were obtained.

\begin{figure}
\centering
\includegraphics[scale=1]{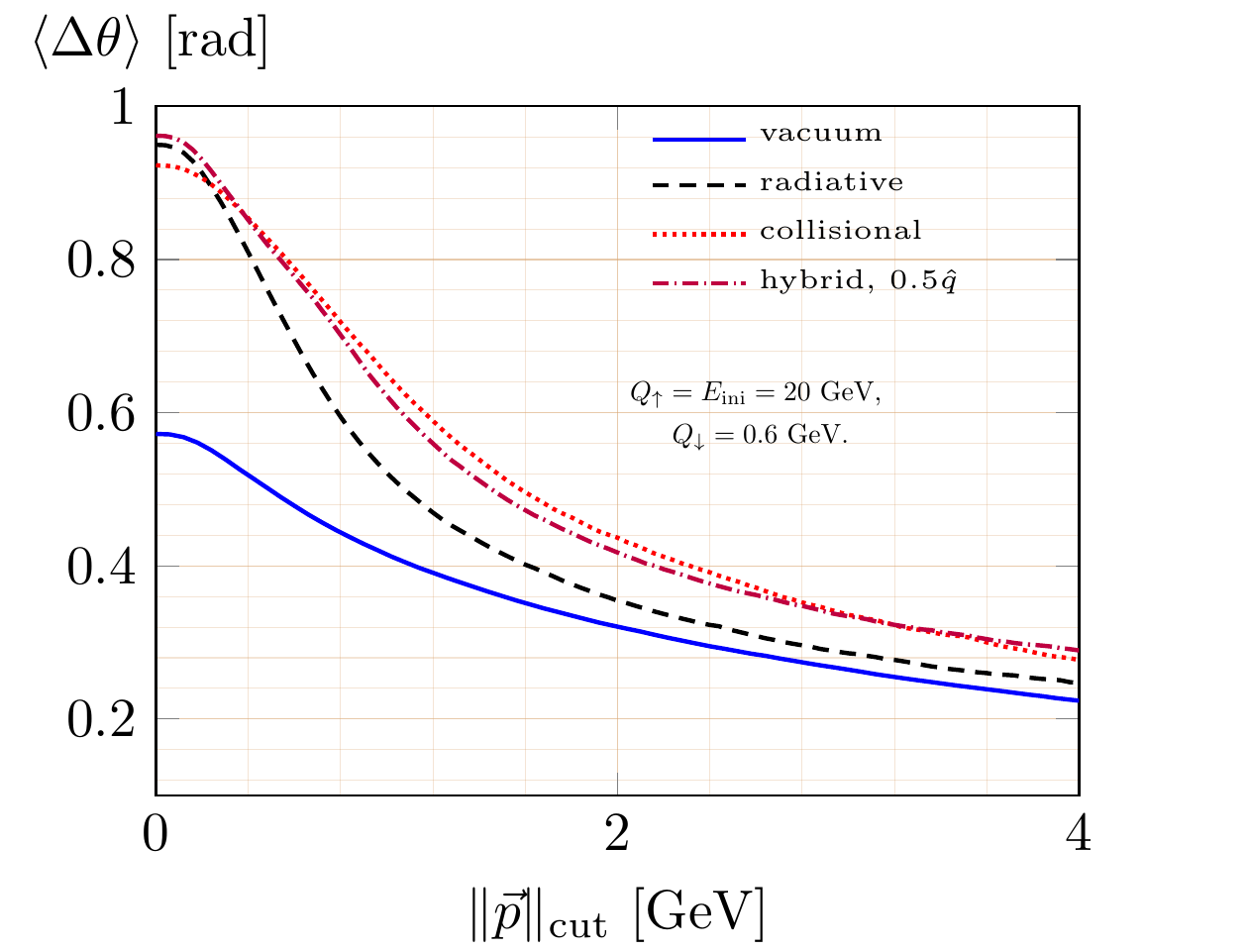}
\caption{Average angle between trigger quark and associated jet-particle as a function of the momentum cut $\|\vec{p}\|_{\rm cut}$ for the radiative, the collisional, and the hybrid model of jet-medium interaction and jets in the vacuum.}
\label{fig:cut}
\end{figure}

For the three models of jet-medium interactions and jets in the vacuum Fig.~\ref{fig:cut} shows some results for the average angle $\langle \Delta\theta\rangle$ as a function of $\|\vec{p}\|_{\rm cut}$, where the average was taken over all simulated jets. For the shown results jets were simulated from an initial quark at virtuality and energy scale $Q_\uparrow=E_{\rm ini}=20$~GeV, down to a threshold virtuality $Q_\downarrow=0.6$~GeV. The initial quark in its final state is then considered as the leading particle of the cascade. In case of the hybrid model, the value of $\hat{q}_C=\hat{q}_R$ was multiplied by a factor of $0.5$, in order to obtain values of energy loss comparable to that of the purely radiative and the purely collisional model.

For the case $\|\vec{p}\|_{\rm cut}=0$, where no jet-particles were rejected, the values of $\langle\Delta \theta\rangle$ are considerably larger in the medium than in the vacuum, while differences between the different models are much smaller. However, towards higher values of $\|\vec{p}\|_{\rm cut}$ the curve for the model of purely radiative energy loss exhibits a much steeper decrease and approaches the curve for the vacuum closer than the two other cases. 
It can be concluded that for the purely radiative model angular broadening occurs due to the radiation of soft particles at large angles. On the other hand, for the purely collisional model and the hybrid model a larger residual broadening can be observed at larger values of $\|\vec{p}\|_{\rm cut}$. 

\section{Conclusion}
We presented a set of effective models for jet-medium interactions, which, although they are rather simplistic, provide a consistent framework to study collisional and radiative energy loss and compare the resulting observables.
 
We implemented these models numerically in a Monte-Carlo algorithm for the simulation of jets.
Among other possible caveats, the current algorithm lacks a hadronization model and a detailed description of the initial hard scattering process. While these restrictions prevent us from quantitative predictions of jet-observables, it is still possible to make qualitative predictions for the behavior of angular correlations: While the model for purely radiative energy loss leads to a broadening that is mainly due to the emission of soft particles at large angles, models with collisional energy loss exhibit a sizeable broadening at larger energy scales due to deflections of the momenta of soft, as well as hard particles. 
\acknowledgments This research was supported by the Polish National Science
Centre Grant\\ No. 2015/19/B/ST2/00937 and by Region Pays de la Loire (France) under contract\\
No. 2015-08473.

\end{document}